# Deployable Payloads with Starbug

Andrew McGrath[*], Roger Haynes

Anglo-Australian Observatory, PO Box 196 Epping 1710 NSW Australia

## ABSTRACT

We explore the range of wide field multi-object instrument concepts taking advantage of the unique capabilities of the Starbug focal plane positioning concept. Advances to familiar instrument concepts, such as fiber positioners and deployable fiber-fed IFUs, are discussed along with image relays and deployable active sensors. We conceive deployable payloads as components of systems more traditionally regarded as part of telescope systems rather than instruments - such as adaptive optics and ADCs. Also presented are some of the opportunities offered by the truly unique capabilities of Starbug, such as microtracking to apply intra-field distortion correction during the course of an observation.

## 1. INTRODUCTION

First described in 2004[1], the Starbug concept allows efficient use of wide field focal surfaces through the use of multiple pickoffs such as fibers, relay mirrors, IFU systems, integrated sensors, etc., deployable freely across the focal surface in parallel. Parallel deployment is a critical element of Starbug, distinguishing it from 'pick and place' robotic positioners such as 2dF[2], OzPoz[3], and others by overcoming the linear dependence of field configuration time on the number of elements to be deployed.

Starbugs are a natural evolutionary development of existing positioner technology. They combine the benefits of traditional "pick and place" systems with those of an Echidna style multi-fiber positioner[4]. The robots are scalable and payload independent – the same design can be used to position single fibers, integral field units, image slicers and pick-off mirrors.

In addition, this technology development opens up other important applications, in particular the development of a reliable positioner for cryogenic/vacuum conditions.

As a comparison, the 2dF robot requires of order one hour to position all 400 fibers on its ~500mm focal surface, an overhead that requires multiple, exchangeable field plates so that a field can be configured 'off telescope', during the period of another observation. An equivalent Starbug-based positioner, even with a bug speed of only 0.5mm/s would be expected to have a typical field configuration time of less than 2 minutes. To change from one essentially random set of bug positions to another, the reconfiguration time actually decreases with increasing bug numbers.

Starbug offers additional benefits over 'pick and place' technologies, of reduced cost, reduced weight and increased reliability due to fewer single-point system failures, and the ability to 'micro-track', or apply small position corrections to the bug positions during the course of an observation, to account for effects such as varying focal surface distortion maps or varying atmospheric refraction due to telescope tracking. Arbitrarily large and arbitrarily curved focal surfaces can be accommodated, enabling wide field multi-object capability on large and extremely large telescopes. Piezoceramic actuator technologies, among others, offer the promise of cryogenic operation for Starbug systems, which constitutes a continuing challenge for pick and place robotic systems. Actuators with promise for Starbug application, based on piezoceramic technology developed for FMOS-Echidna, have been demonstrated at the AAO[5], underpinning the feasibility of these concepts.

In the following sections of this paper, we explore some of the parameter space for instrumentation concepts that take advantage of the unique capabilities of Starbug-type actuators for positioning of multiple payloads in the focal surface.

*ajm@aao.gov.au; http://www.aao.gov.au; phone +61 2 9372 4848; fax +61 2 9372 4880

## 2. REFORMATTING THE FOCAL PLANE

There are a range of applications involving positioning technologies in telescope focal surfaces. Essentially, they are all driven by the need to make efficient use of detector real estate. Wide field astronomy, defined in terms of the number of resolution elements spanning the telescope's corrected field of view rather than any specific physical angle on the sky, is a growing field. The increasing success and planned implementation of adaptive optics on large and extremely large telescopes yields ever-increasing numbers of resolution elements. Instrument cost increases in proportion with this effect, to provide detectors for imaging and/or spectroscopy of all the available field.

Wide field imaging or spectrographic systems (FMOS, 2dF, FLAMES, 6dF, SuPrimeCam, and proposed instruments such as WFMOS, HyperSuPrimeCam and MOMSI[6]) operate in corrected focal surfaces that are very large in terms of numbers of resolution elements.

In general, most science undertaken by these wide field systems only utilizes a small fraction of the available field, with objects of interest separated by relatively large uninteresting regions (in essence, the sky is mostly black, with small targets). An obvious strategy to maximize the effectiveness of investment in instrumentation is to limit data collection to the interesting regions – to rearrange the light from the focal surface in such a way as to put just the interesting bits onto the detectors. This logic has led to several classes of instrument now familiar to us – fiber-fed multiobject spectrographs, image slicers, integral field units, and more.

The Starbug paradigm enables instrument concepts to extend to design spaces that have till now been inaccessible or at least highly challenging to previously implemented focal plane positioning technologies. In particular: vacuum and cryogenic operation, very large and arbitrarily curved focal surfaces, intelligent payloads, and continuous or mid-observation field reconfiguration all become much easier to implement.

## 3. ACTUATOR REGIMES

Depending on the specific application, Starbug encompasses a wide range of microrobotic actuators, with different characteristics. Bugs may carry active or passive payloads. Bugs may operate wirelessly or with service umbilicals. In some applications with high payload densities and/or tightly constrained focal surface sizes, a small footprint is important, but for fewer, larger payloads a quite different underlying bug technology may be most appropriate, taking advantage of the greater space available. Focal surface positioners also have a broad spectrum of positioning accuracy requirements, depending on the plate scale, imaging resolution and science application.

It is not the purpose of this paper to explore the specific technologies that may be employed in such a broad range of possible bug designs, but we point out that any given bug technology is unlikely to have optimum characteristics for all Starbug applications, and that a deficiency in capability on one area for a targeted bug development may well not be a disadvantage for another application.

## 4. STARBUG ADVANTAGES

### 4.1. Microtracking

A particular and unique capability offered by Starbug is the possibility of microtracking and related 'field tweaking' concepts. Here, the field configuration (i.e. the arrangement of positioned elements in the focal surface) is adjusted during the course of an observation.

During the course of an observation, the telescope tracks the target field as is moves through differing elevations. Atmospheric refraction causes a variation in plate scale across the field through varying airmass depth due to this effect, and so objects within the field move with respect to each other during the observation. The magnitude of this effect for relatively large field (tens of arcminutes) can be huge. When observing at high air masses the physical target separation at the extremes of the field can change rapidly. For example, with the 30 arcminute field of FMOS observing at high air masses, the rate of target motion at the extremes of the field can be in excess of 0.5arcsec/hour. For 2dF on the AAT

(with its 2º field of view) this it can be >2arcsec/hour. This sets a very hard upper limit on the maximum observing times for a particular field configuration before a configuration tweak is required. For both FMOS and 2dF ideally this is less than 30 minutes when observing through a rapidly changing air mass. If there is a requirement to work at moderate to small spatial scales, even over relatively moderate field view field, it is vital to consider the impact of this field distortion. Regular field tweaks or continuous microtracking may be essential if integration times of more than a few minutes are required.

Further to the distortion issues there is also the chromatic dispersion of the atmosphere. This again can be at arcsecond scales, however, this can be largely corrected with an ADC if the system design allows. However even here there can be a small residual differential dispersion across the field. This residual differential dispersion only really starts to have an impact at very fine spatial scales. Calculations of differential refraction show that for a zenith angle change from 45 to 60 degrees, the atmospheric refraction effects distort the plate scale such that targets move approximately 20 milliarcseconds with respect to each other across a 1-arcminute field-of-view[7]. Proposed ELTs have specifications approaching these limits, and so a microtracking capability, or at least short-term reconfigurability, are essential to preserve optimal performance for MOS instruments on telescopes in this class.

The potential to move bugs during an observation also allows for preparation for the next field. In a case where a significant excess of Starbugs are implemented beyond the science requirements for a single exposure, it would also be possible to imagine repositioning a subset of (unused) actuators during the course of an observation, in preparation for the next observed field.

### 4.2. Field configuration time

Simultaneous activation of actuators inherently reduces the time to position a set of Starbugs for an observed field. To first order, the configuration time is now independent of the number of payloads to be positioned. This is similar to the field configuration time capabilities of FMOS-Echidna, and compares favorably with pick-and-place technologies such as 2dF, where configuration time is proportional to the number of payloads. We note here that FMOS-Echidna has a field configuration time of ~10 minutes for 400 fibers, but that in this case the bulk of the configuration time is spent on fiber metrology for position feedback, rather than in actually moving the fibers.

Even a 2 meter field plate with 400 bugs can typically change from one (effectively random) configuration to another in 10 minutes with a bug speed of less than 0.5mm/s. Prototype bugs using resonantly-enhanced piezoelectric actuators have already been demonstrated operating at speeds more than an order of magnitude in excess of this[5].

### 4.3. Reduced size and mass

Positioning robots using existing technology are normally of substantial mass. This is because of the required mechanism precision and consequent need for stiffness and stability. The current and next generation of large and extremely large telescopes tend to have physically large focal surfaces; the VLT Nasmyth focus is corrected over a field that results in a focal surface approaching a meter in diameter. The reflective Schmidt-type LAMOST has a focal surface 1.2 meters across. OzPoz, the pick-and-place fiber positioner for FLAMES operating at the Nasmyth focus of the VLT[3], has a robotic positioning gantry and a rotary plate exchanger that is several meters high, weighing near 2000 kg and consuming a substantial fraction of the available mass budget for that focal station (Fig. 1).

The Starbug concept decouples the size of the robotic components from the size of the focal surface. Actuators are small, quite independently of the size of the focal surface. A positioner operating with a 2-metre focal surface is little more difficult or costly to implement than on a 300-millimetre one. A field plate is fitted to the focal station, with little overhead requirement. It only needs to carry the Starbug actuators, controlled by an electronics support system that may be mounted separately.

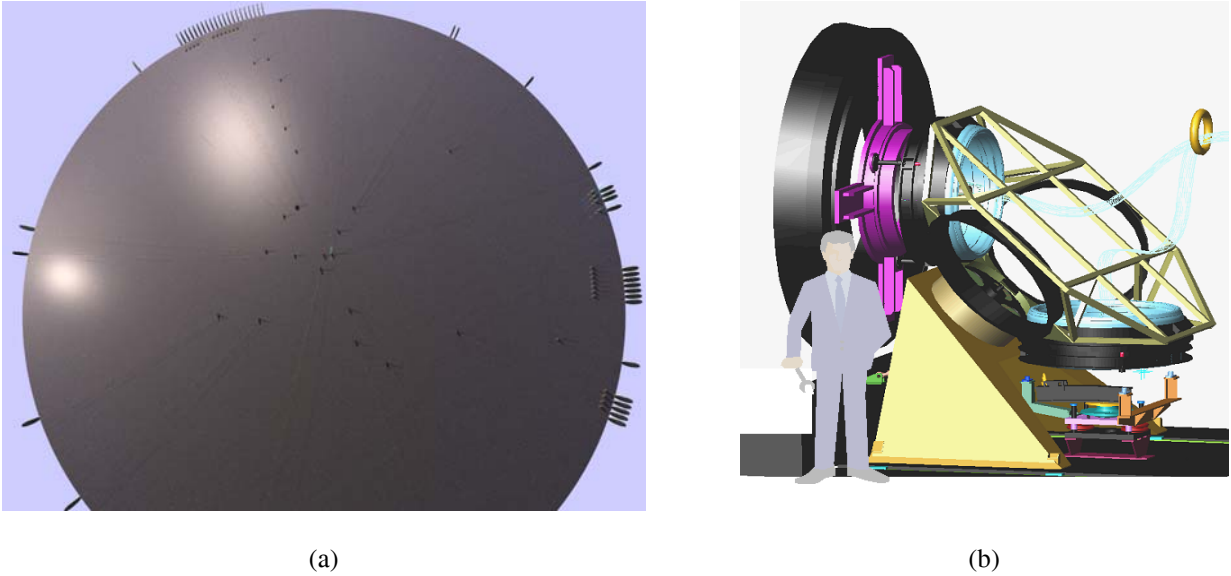

Fig. 1. (a) The Starbug concept allows payload positioning across an arbitrarily large field plate. Here, a 2-m diameter plate is partially populated with Starbugs carrying fibers for a multi-object spectrographic system, with the fibers and actuator electrical feeds managed by retractors similar to those implemented for pick-and-place fiber positioners. Note the lack of requirement for heavy, large robotic systems, such as that of OzPoz, the positioner for the FLAMES facility on the VLT (b).

### 4.4. Cryo accessibility

The micro-robotic actuators developed for FMOS-Echidna incorporate no lubricants, are constructed from vacuum-friendly components, and can operate at low temperatures (albeit with reduced speed approximately proportional to absolute temperature). With robotic actuators based on the same technology, a Starbug system can have cooled or even cryogenic capability. Development of the technology is likely to provide a variety of cryogenic-capable mechanisms, which experience shows can often be problematic in the development of astronomical instrumentation.

Many attractive options for instrument concepts involve focal plane positioning systems requiring cooled or cryogenic focal surfaces. K-band observations suffer greatly from the thermal emission of components in or near the optical path. The capability of certain common microrobotic technologies, such as piezoelectric actuators, to operate at greatly reduced temperatures enables a range of positioners to be imagined, operating in cryostats for reasons of minimizing thermal emission or reducing noise in active payload components such as sub-field imagers.

### 4.5. Scalability

As stated previously, Starbug decouples the scale of the robotic components from the scale of the focal surface. Largely because of this decoupling, a Starbug system can be arbitrarily large, limited by the size of the corrected telescope field or other infrastructure (such as refrigerated or cryogenic enclosure limitations).

The Starbug multiplex capability lends itself strongly to scaling. Until bug density limitations are encountered, the number of bugs can be varied easily, with the system cost scaling close to linearly with the number of bugs.

### 4.6. Redundancy

An important aspect of the Starbug concept is that the actuators are largely independent. Pick-and-place positioners are vulnerable to a variety of single-point failures, where failure of a single component renders the system out of service. The distributed nature of the Starbug robotics means that many failures will only affect a single actuator. The system degrades much more gracefully, with reduced performance in response to many component failures, rather than total system incapacity.

Further, it is likely that the actuators can be very simple, reliable and long lifetime mechanisms based on a very few piezoelectric ceramic components. Their independent operation reduces the number of possible single point failures that disable the entire system.

### 4.7. Instrument upgrade path

A system based on Starbug concepts lends itself to future upgrades by its modular nature. Once a Starbug paradigm has been adopted, different (or more of the same) Starbug system components can be added without disturbing the fundamental architecture. A much high degree of planned upgrading and future-proofing is thereby achieved compared with systems that tightly integrate important system functionality with major physical structure.

In particular, bugs carrying active sensing payloads of new and different types can be readily added at any time in the instrument's life – it's an ideal prototyping environment, while remaining a facility class instrument.

## 5. INSTRUMENT CLASS APPLICATIONS

We conceive a range of classes of instrument concepts that would be enabled or facilitated by Starbug-type positioning technologies. These range from making it easier to build instruments with capabilities similar to existing facilities, through to concepts that critically depend on unique Starbug characteristics.

### 5.1. Fiber-fed, discrete object Multi-Object Spectroscopy

Coming from the perspective of the current generations of robotic focal plane fiber positioners (epitomized by 2dF and FMOS-Echidna), this is perhaps the most obvious application of Starbug technology. In instruments of this class, Starbug actuators patrol a focal surface with more or less freedom of motion, each carrying a single optical fiber, somewhat like 2dF but where the magnetic buttons can be independently and simultaneously moved without the need for a large and precise robotic mechanism (Fig. 2). In this application, an optical fiber already forms a 'tether' for each bug, and thus provides an obvious route for service feeds (power and control) to the bugs.

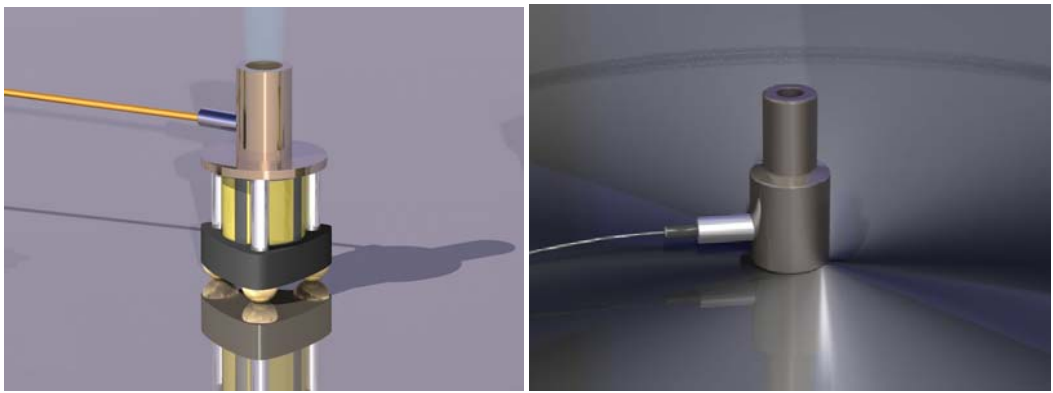

Fig. 2. A Starbug implementation (left) of a discrete-object fiber positioner uses magnetic buttons similar to those carrying the fibers for OzPoz (right) and other pick-and-place positioners, but mounted on micro-robotic actuators that can be independently and simultaneously positioned by 'walking' across the field plate.

Discrete MOS Starbug instruments offer clear and dramatic weight savings over pick-and-place technologies when the focal surface is large. A single focal plate suffices because of the relatively short configuration time that results from simultaneous bug motion. Cooled and even cryogenic MOS applications are far easier to consider, and the concept can readily accommodate microtracking in an era of sub-seeing telescope imaging performance.

### 5.2. Deployable IFU Spectroscopy

Increasing use has been made in recent years of the 3-D datacubes available from integral field unit-type data. In the case of fiber fed instruments, microlens-fed integral field units have typically been fixed in the focal surface, such as Argus in FLAMES on the VLT, or CIRPASS, built by the Institute of Astronomy at Cambridge. Other instruments achieve their functionality with image slicers, also fixed in the focal surface. Cases have also been built for deployable IFUs: the FLAMES facility provides 15 deployable small (20 element) IFUs, and KMOS[8] is already under construction using a complex set of mechanisms to achieve this functionality in a cryogenic environment.

Clearly, deployable IFU pickoffs (dIFUps) on the FLAMES model can be readily accommodated by Starbug in much the same manner as discrete object MOS. However, the nature of Starbug movement opens up a much broader horizon for IFU configuration. Starbug is a very plausible mechanism for implementation of tileable IFU arrays, as proposed by Bland-Hawthorn *et al.*[9] In that vision, large numbers of small deployable IFU units may be arranged to form contiguous or discontinuous 'meta-IFUs' of arbitrary shape (Fig. 3), to take maximum advantage of limited detector real estate by limiting the observational focus much more precisely to the regions of interest.

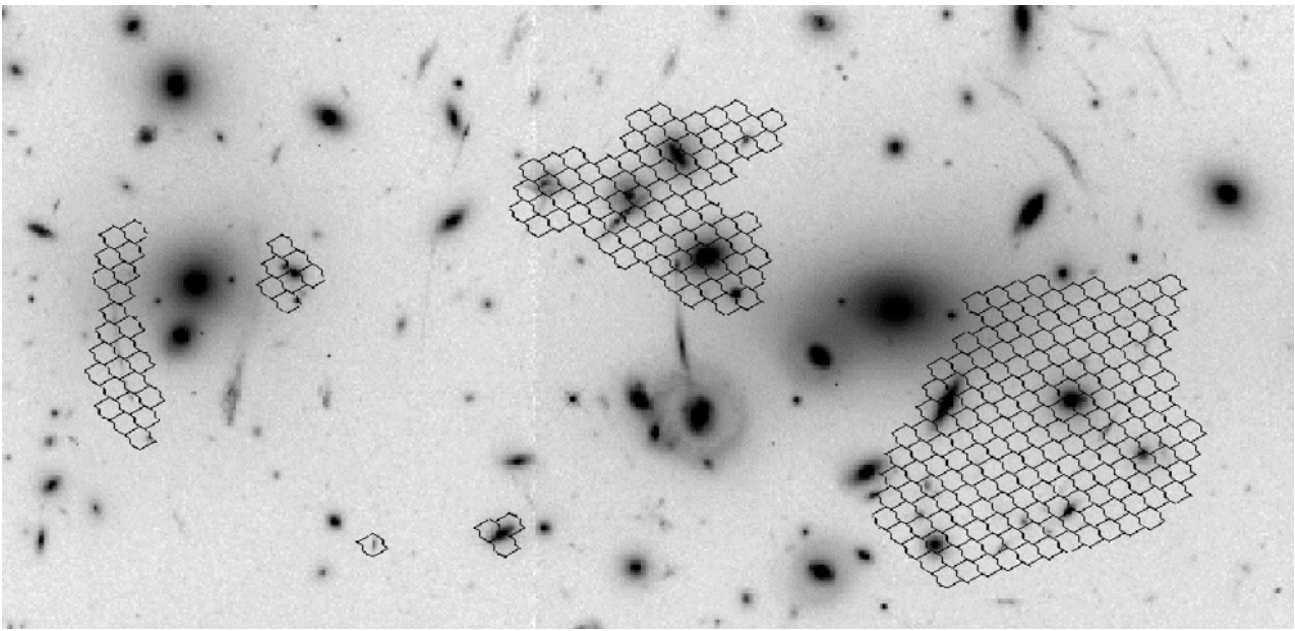

Fig. 3. Honeycomb configuration of the strongly lensed galaxy cluster Abell 2218 (HST image). The field has been configured showing the advantages of small but tileable deployable IFU sub-units. Figure from Bland-Hawthorn et al. (2004)[9].

### 5.3. Subfield correction

Traditionally in astronomical equipment, there has been a clear divide between telescope and instrument. The telescope delivers a corrected focal surface, and the instrument samples or measures it in various ways. Adaptive optics is widely recognized as truly critical to the success of the next generation of telescopes, where it needs to perform successfully at a high level in order to deliver the potential capabilities of these extremely large telescopes.

Unfortunately, the challenges of adaptive optics typically become more difficult with increasing telescope size. In particular for wide-field astronomy, the challenges of adaptive optics become more difficult with increasing telescope field of view. Typical modern 'wide field' AO systems only operate effectively (Strehl of a few tens of percent) over fields of view of one to two arcminutes – far from 'wide field' in terms of multi-object spectroscopy, for example.

Adaptive optics systems take a variety of forms and involve complex systems operating at various points in the optical train. The FALCON spectrograph concept[10] distributes the application of correction, with first order correction applied globally across the entire field, and higher order correction applied locally across a subfield. This reduces the otherwise intractable challenge of achieving high levels of correction across wide fields. The Starbug notion offers the possibility of using active components located in the focal surface to apply such partial correction locally to subfields.

In a related manner, Starbug-based local correction may undertake second-order (residual) correction for active optics (aO) or chromatic atmospheric dispersion compensator (ADC) errors, easing the design requirements on these other parts of the telescope corrective optics which now only need to manage first order effects. The basic philosophy seen in this approach is that when wide-field correction of any imaging distortion effect is difficult, then local correction of discrete subfields may be a technically more feasible, or simply more cost-effective, solution – sharing the problem between low-order wide field correction, and higher-order narrow field correction.

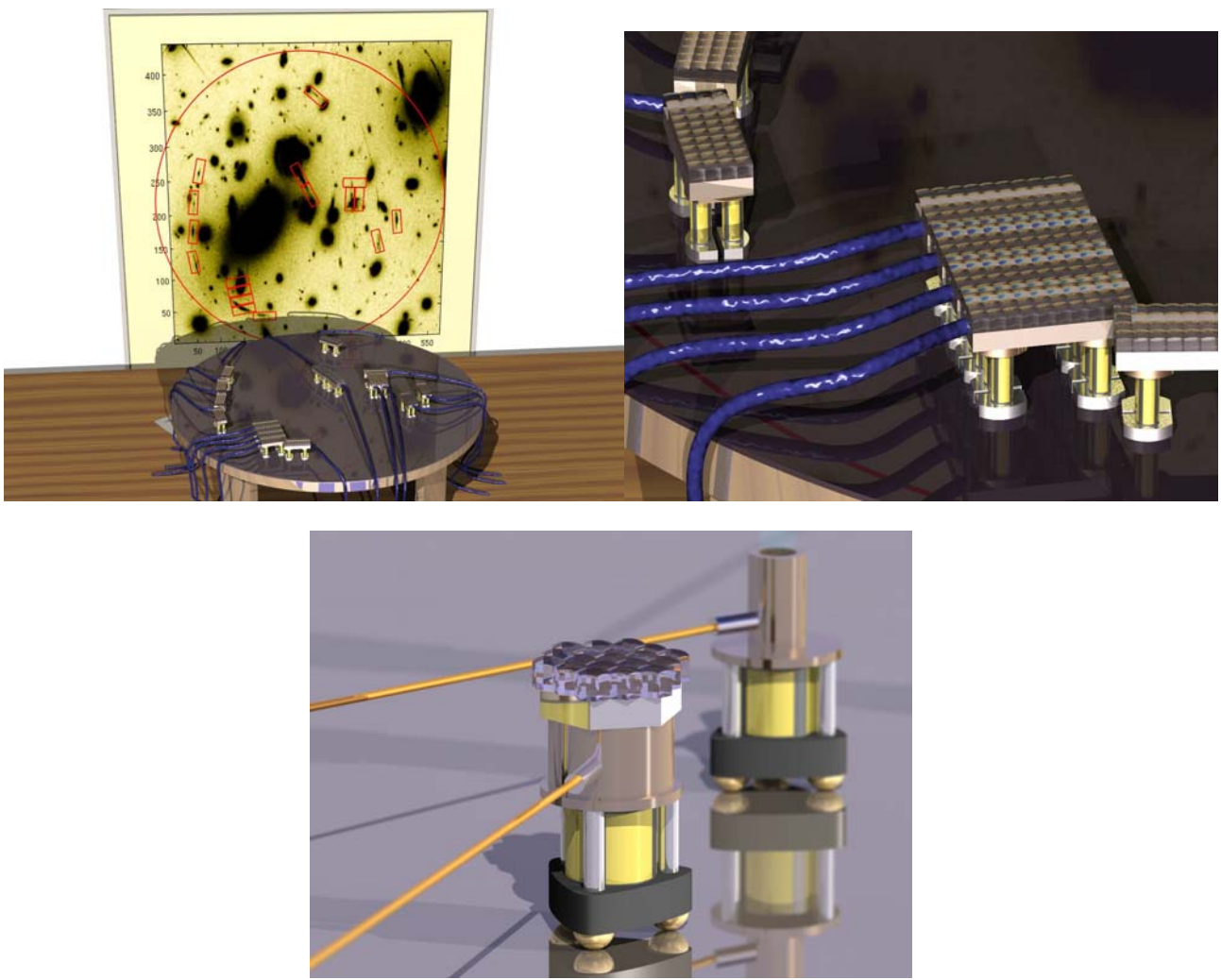

Fig. 4. Two possible implementations of tileable or edge-buttable deployable IFU pickoffs. Rectangular deployable units (above) may well-suit certain types of target, while hexagonal or other continuously tileable shapes (below) may better accommodate truly arbitrary regions-of-interest.

### 5.4. Subfield imaging

Analagous to the multiplex advantage inherent in multi-object spectroscopy, sub-field imaging allows effective use of detector real estate by only collecting data over regions of interest. A Starbug-based sub-field imaging system could operate in a variety of ways.

It could carry pickoff mirrors, lenses and/or prisms to relay a localized segment of the field to an imaging camera, generically illustrated in Fig. 5. Such an implementation could well incorporate some measure of payload manipulation (e.g. tip/tilt pointing control) to align the subfield imaging optics, as suggested in Fig. 6. Such an optical relay solution lends itself to wireless bug technologies, since there is no payload function requiring service feeds. A complexity associated with this relay concept is that path lengths could be high, leading to a large pupil divergence. If pupil control is also mandated, a path length compensator will be needed to accommodate the different path lengths to different bug positions. This could be achieved in the optical system off of the field plate, perhaps using a traditional trombone-style path length compensator, but an interesting option may be to handle the length adjustment with Starbugs. In such a solution, the light is not relayed directly from the 'receiving' bugs to the collection optics, but rather via a second bug, carrying a fold mirror to manage the overall path length.

Clearly, at all telescope foci other than extremely slow beam adaptive optics systems (f/30 or more), the receiving bugs should carry more than a flat mirror to pick out the desired subfield, because of the beam divergence. Even if the beam is collimated by the bug optics, a non-zero field of view means that the pupil will diverge. It is likely that a slow beam (non-collimated) will lead to the best compromise on beam and relay optics diameter.

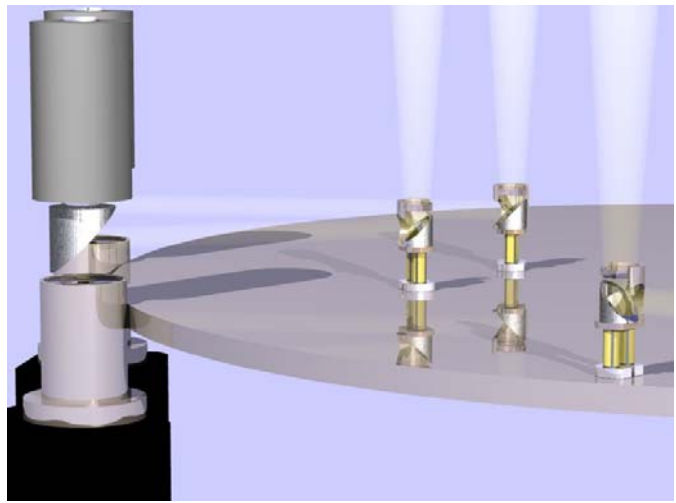

Fig. 5. Generic concept of a Starbug-based optical relay for subfield imaging.

Another arrangement to provide subfield imaging would be to carry a coherent imaging fiber bundle, much as a high-spatial-resolution IFU. This is likely the simplest subfield imaging solution, technically similar to a deployable IFU instrument concept. A bundle could take the form of a traditional assembly of conventional fibers, as is commonly implemented for IFU instruments, but it may also prove effective to use an integrated, multi-core fiber for reasons of reduced bulk and finer spatial resolution[11]. Such multi-core fibers with thousands of discrete, isolated cores, are already commercially available[12]. As an area of active commercial development, we can expect the throughout, crosstalk, FRD and bandwidth performance of such devices to increase.

Alternatively and more directly, a subfield imager could use Starbug actuators to position bare detectors in the focal surface. In such a configuration, we have effectively shifted the Starbug concept into an imaging camera to give a reconfigurable detector array. Optimum CCD performance requires detector temperatures in the order of 150 to 200K, and piezoelectric Starbug technologies certainly may operate in this regime. Filter exchangers would be a substantial complication to such an instrument, however a simple solution is available when the system is implemented with a number of bugs. In this case, each bug need only carry a single filter, with a range of filters offered by a range of bug 'flavours'. Any given observation may then view different targets through different filters simultaneously, which are then reconfigured for subsequent observations of the same field to accumulate all the required filter exposures. Another solution may be the emerging vertically integrated array technology[13], offering a convenient method of electrically switching between detector sensitivity wavebands – effectively selecting different filters.

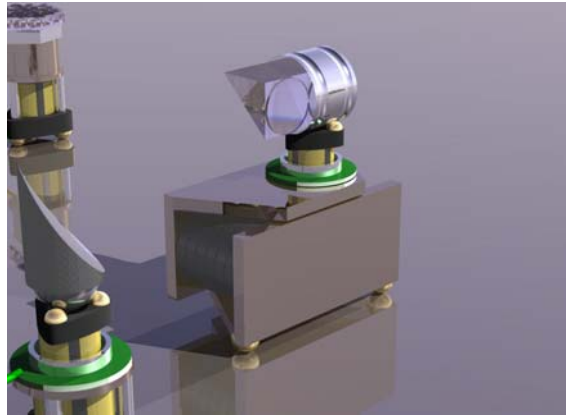

Fig. 6. Prism-based sub-field pickoff optics carried on a Starbug for subfield imaging. This version incorporates a payload manipulator providing elevation angle control of a prism/lens assembly

### 5.5. Relayed Image Multi-Object Spectroscopy

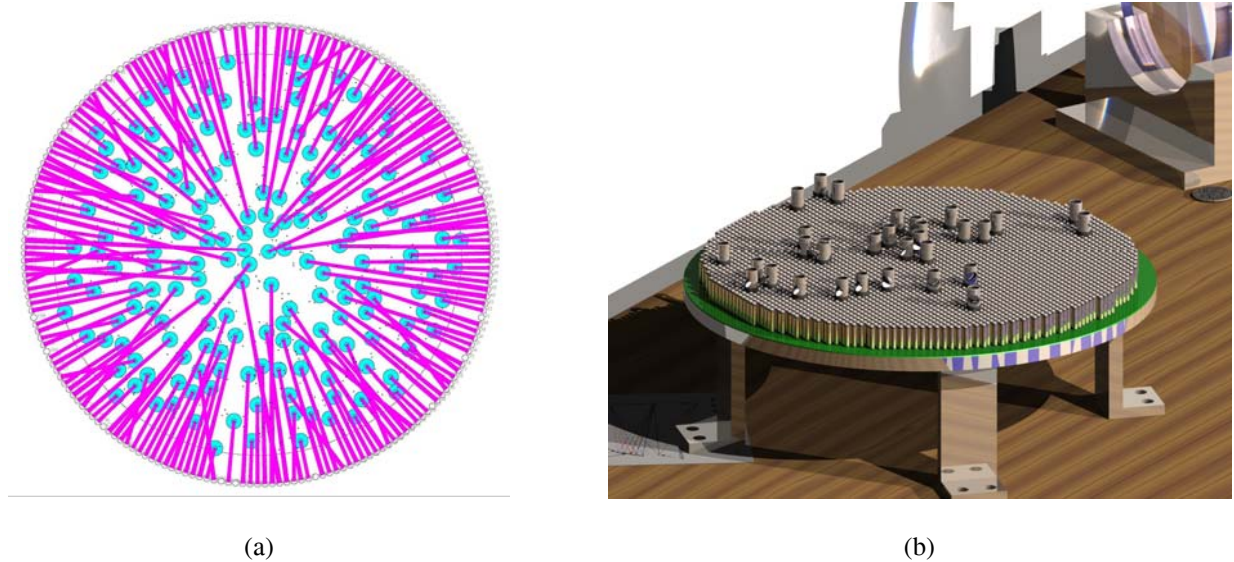

(a) (b)

Fig. 7. (a) Simulated Starbug image relay MOS field configuration, showing 178 out of 200 bugs allocated to targets from 800 randomly distributed targets on a 300-mm diameter focal surface. The bugs have a 10-mm footprint, and pick off a subfield 3 mm in diameter (figure from Bailey, 2004[7]). (b) shows a implantation of this – only 36 bugs are shown for clarity. The bugs in this version are passive, positioned using a robotically active surface as a field plate.

In many ways, this concept is equivalent to conventional image-sliced IFU spectroscopy, but the Starbug aspect adds arbitrary reconfiguration of the slices, giving access to discontiguous subfields. An instrument on this model has already been seriously proposed, MOMSI, a MOS instrument concept investigated for 30-100m telescopes.

For exactly the same reasons as a an optical relay-based subfield imager, this type of instrument will need to include a path length compensator to deal with the different path lengths from various bug positions to the spectrograph optics. Again, this could be accomplished using a trombone-style adjustment as proposed for MOMSI[7], or by using additional Starbugs carrying simple fold mirrors as part of the relay optical path before the light even leaves the field plate.

### 5.6. Active science payloads

The Starbug concept is particularly powerful when combined with modern advances in miniaturized sensing technologies. A very direct model for sampling the focal surface is to carry the sensors themselves physically to the regions of interest. This approach has already been discussed with respect to subfield imaging, however a wide range of different sensor types can be used in this way.

Simple light detectors (avalanche photodiodes, photomultipliers, or other) can perform photometry without introducing additional noise and uncertainty by transmission through unnecessary optical elements. In its most direct form, such photometers could be placed directly at prime focus, with just a single reflection from the telescope primary in between the starlight and the detectors. These could provide highly time-resolved output, revealing rapid occultations or other variability, or integrate for long periods.

Recent advances in holographic optical element design and photonics also raises the possibility of deployable microspectrographs. Throughput for such an arrangement is likely to be particularly high, since this arrangement combines fiber-like field configurability with slit-fed spectrographs. The Georgia Institute of Technology has recently announced development of an inexpensive, very small spectrograph[14] just tens of millimeters in scale (Fig. 8), based on a volume phase holographic grating with optical power so it can also serve as an imaging optical element of the spectrograph. Alternative dispersers enabling microspectrographs are also arising from recent developments in photonic crystal superprisms[15]. The AAO has also been involved in exciting work developing an integrated photonic spectrograph for astronomical application[16].

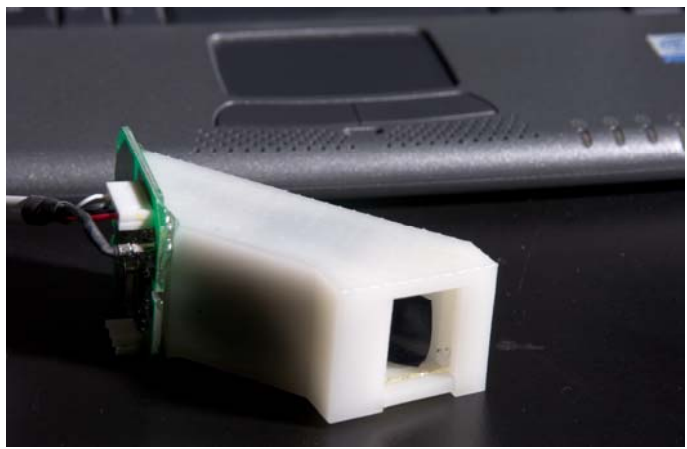

Fig. 8. The Georgia Institute of Technology has created this tiny and inexpensive spectrograph, just 60mm long and suitable for even smaller implementation. Such developments raise the possibility of an instrument based on deployable microspectrographs.

Tunable filter payloads are also an interesting possibility, especially when combined with deployable imagers. An acousto-optical filter, or a fabry-perot tunable etalon, may be incorporated to a Starbug payload to provide this functionality. Recent developments involving integrating a tunable filter directly onto the surface of a detector array, also promise compact and lightweight tunable filter imaging[17]. Current technology in STJ arrays falls short of the datacube collection capabilities of other presently available technologies, but future developments may make such direct photon energy sensing arrays useful as deployable Starbug payloads.

### 5.7. Active Payloads supporting Telescope Functionality

Apart from subfield imaging by carrying bare detectors in the focal surface, other active, imaging payloads may be used for telescope support purposes. In particular, relatively basic types of detectors carried in a similar way may be used as guide probes. Lacking the need for the sensitivity of science detectors, such a guide camera may be suitable for room temperature (dome temperature) operation, and so a Starbug-based guide probe may coexist with other room-temperature Starbug science payloads, without significant additional infrastructure.

In a similar way, on-instrument wavefront sensors may be carried on Starbug actuators providing feedback for active mirror support systems to maintain the telescope primary's figure.

Used for these purposes, relatively large Starbug actuators are likely to suffice, easing implementation challenges and giving significant capacity for carrying the more complex payload service feeds.

### 5.8. Simultaneous observing with mixed payload types

Starbug technology offers another degree of freedom to the instrument designer – the payloads of the bugs no longer need to conform to the same degree of standardization required for the pick-and-place gripper. This enables a variety of payloads to be positioned within a single instrument. Within bug population density limitations, effectively multiple instruments can be implemented, able to access the sky simultaneously by sharing the focal surface.

Although clearly such a shared telescope use imposes observing restrictions (of exposure time, and the need to choose different objects within the same telescope field), this may be a useful mechanism for effective utilization of telescope time. It could also serve simply as a convenient mechanism for instrument exchange, with one set of bugs (say, dIFUps) retiring to parked positions to allow sky access for another set (say, subfield imaging).

A precedent for such a "multi-instrument instrument" already exists in the form of FLAMES on the VLT. Here, discrete object fibers, deployable IFU pickoffs, and a fixed larger format IFU all coexist in a single instrument. Another example is the proposed WFMOS facility intended for the Subaru telescope[18]. Here, an Echidna-type positioner allocates some of its fibers to high resolution spectrographs, targeting a science case based on a survey of stellar metallicities, and some to low resolution spectrographs, targeting a science case based on a galaxy redshift survey. The differently-allocated fibers are spatially interleaved and allow simultaneous observation of both surveys, saving substantial survey time.

## 6. CONCLUSIONS

Overcoming many of the limitations of both pick-and-place and Echidna-type robotic positioners, Starbug is an exciting concept for robotic positioning technology. This concept offers the potential to open a new instrument design space, enabling conception of new classes of instruments that reformat the focal plane to focus only on those areas of interest in a wide, and hence expensive to fully sample, corrected field.

The flexibility of the Starbug concept may be used for other focal surface positioning applications, such as pickoff mirrors for an image relay or even deployable micro-imagers or micro-spectrographs. Further, the technologies involved in Starbug are particularly well suited to cooled, cryogenic and vacuum environments, and so instruments requiring positioning robots in these situations may be considered.